\documentclass[amsmath,amssymb,prb,preprint]{revtex4}

\usepackage{graphicx}
\begin{document}

\title{Single-mode phonon transmission in symmetry broken carbon nanotubes}

\author{Jian Wang$^{1}$}
\author{Jian-Sheng Wang$^{2}$}

\affiliation{$^{1}$College of Physical Science and Technology,
Yangzhou University, Yangzhou 225002, P. R. China \\ $^{2}$Center
for Computational Science and Engineering and Department of Physics,
National University of Singapore, Singapore 117542, Republic of
Singapore }

\date{25 July 2008}

\begin{abstract}

Normal mode phonon transmissions are studied in carbon nanotubes
with the presence of Stone-Wales (SW) defect, using a mode-matching
method and through the analysis of symmetry. The calculation shows
that the transmission for low group velocity acoustic phonons is
evidently reduced at high frequency range, and that this SW defect
induced symmetry breaking  strongly inhibits the transmission of
long wave optical phonons in carbon nanotubes. The characteristic
features of
 transmission for each phonon mode depend on the symmetry. These
 findings suggest that the local heating in the defective nanotubes
 may be contributed mainly from the low group velocity acoustic phonons and optical phonons near the $\Gamma$-point.

\end{abstract}


\maketitle Carbon nanotubes (CNTs) are uniquely
symmetric\cite{symvic,symbook,sysalon} quasi-one-dimensional
structures with remarkable electrical, mechanical, and thermal
properties for various engineering applications. \cite{symbook,
nanotubebook,nanotubereview,ballistic1,ballistic2} However, such
symmetries for perfect carbon nanotubes are usually
broken\cite{nanotubereview}, either for the
 purposes of engineering application such as the junction structure,  or
due to the technical flaws during the process of synthesis such as
the Stone-Wales (SW) defects. When the symmetry of structures is
broken, the equation of motion that underlines the physical
properties of material does not respect the symmetry any
more.\cite{anderson} Therefore,  such symmetry breaking in nanotubes
is important for deep insight into both physical properties and
transport behavior. The total transmissions for all possible normal
mode phonons at a given frequency have  been first studied using the
nonequilibrium Green's function method (NEGF).\cite{negfyamato} Of
particular interest in this letter is the single-mode phonon
transmission of CNTs in the presence of the SW defect, which is
known as the explicit symmetry breaking mechanism. Understanding
single-mode phonon transmission in the symmetry broken nanotubes is
not only fundamental to addressing the heat dissipation in
nanodevices which challenges the current electronic miniaturization
and the thermal transport in carbon
nanotubes,\cite{nanotubebook,nanotubereview,ballistic1,ballistic2}
but also relevant to the phonon control problem in the emerging
phononic devices.\cite{phononcontrol,phononcontro2,phononcontro3}
This letter aims to study the behavior of single-mode phonon
transmission in the symmetry broken nanotubes with the SW defect
through the analysis of its symmetrical property. We first introduce
the mode-matching method\cite{modecoupling,generalvalue,reviewus}
used in our calculations. Then the phonon dispersion of the
single-walled achiral carbon nanotube $(11,0)$ is calculated from
the force constants derived from the second-generation Brenner
potential.\cite{dwbrenner} Each branch of normal mode phonon is
further analyzed within  the full symmetry of line group theory.
Finally we present the calculated results of single-mode phonon
transmission in the symmetry broken CNT$(11,0)$ in the presence of
the SW defect, using the mode-matching method.

The NEGF\cite{negfus,negfyamato,ngfmingo,reviewus} has been employed
to calculate phonon transmission coefficient which is the sum of all
the normal modes from the leads. It is  in principle at least
applicable  to both the elastic and nonlinear scattering
regime.\cite{reviewus} However, it is difficult to analyze the
contribution from each mode, respectively. The mode-matching
method\cite{modecoupling,generalvalue,reviewus} serves as another
equivalent way to calculate the transmission coefficient in the
ballistic limit. In comparison with NEGF, the most prominent advantage of the mode-matching
method is that the single mode phonon transmission can be obtained
from this method explicitly. But the mode-matching method is  valid
only in the elastic regime, where the anharmonic scattering is
neglected. In the low-dimensional carbon nanotubes, anharmonic
scattering does not play an important role in thermal transport
below room temperature and thermal transport is found almost
ballistic through the experiments.\cite{ballistic1,ballistic2} The dominant
scattering mechanism is elastic due to topological defects or
impurities in CNTs at low temperatures. Therefore, we focus our discussion  on the elastic
regime. The mode-matching  is the exact method of calculating the
single mode phonon transmission for the elastic scattering.

 We assume that one normal mode\cite{modecoupling,generalvalue,reviewus}
$ \tilde {\bf u}_{l,n}(\omega, {\bf q})=\tilde {\bf e}_n  e^{i({\bf
q}\cdot {\bf R}_l-\omega t)}$
 is incident from the left lead, where each eigenvector $\tilde {\bf e}_{n}$ satisfies the dynamic
equation of motion\cite{reviewus} ${\bf D}\tilde {\bf e}_{n}=\omega
^2 \tilde  {\bf e}_{n}$. Here ${\bf D}$ is the dynamic matrix for
the unit cell on the perfect lead, $l$ denotes the atom in the $l$
unit cell, ${\bf R}_l$ denotes the position for the $l$ unit cell,
and $n$ refers to the polarized phonon branch.  Angular frequency
is denoted as $\omega$, wavevector as $\bf q$. Due to the defect
scattering in the central region, the solution for the left/right
perfect leads can be written as\cite{reviewus}
\begin{subequations}
\label{boudaryeq}
\begin{eqnarray}
{\bf u}_{l}^{L} &= & \tilde {\bf u}_{l,n}(\omega, {\bf q}) + \sum_{n'}t^{LL}_{n'n} \tilde {\bf u}_{l,n'}(\omega, {\bf q'}), \\
{\bf u}_{l}^{R}&=& \sum_{n''}t^{RL}_{n''n} \tilde {\bf
u}_{l,n''}(\omega, {\bf q}''),
\end{eqnarray}
\end{subequations}
where $n, n', n''$ refer to  the different polarized branches
of incident, reflected and transmitted  waves. Wavevectors  for the
incident, reflected, and transmitted waves are ${\bf q}, {\bf q}', $
and ${\bf q}''$, respectively.  The superscript $L$ and $R$ indicate
the left and the right.
 In these equations, $t^{RL}_{n''n}$, $t^{LL}_{n'n}$ are
the amplitude transmission/reflection coefficients from mode $n$ on
the lead $ L $ to mode $n''$ on lead $R$ and to mode $n'$ on lead
$L$. The wave vectors $\bf q'$ and ${\bf q}''$ are found to satisfy
$\omega=\omega_{n'}({\bf q'}) = \omega_{n''}({\bf q}'')$ using the
generalized eigenvalue decomposition method.\cite{generalvalue} Note
that frequency does not change because the scattering is elastic.
The group velocities $v_g$ for the reflected waves $\tilde {\bf
u}_{l,n'}(\omega, {\bf q'})$ are backward $v_g<0$, for the
transmitted waves $v_g>0$. In order to obtain the coefficients
$t^{RL}_{n''n}$ and $t^{LL}_{n'n}$, an alternative, but equivalent,
form of expression of Eq.~(\ref{boudaryeq}) has been proposed in
Ref.~\onlinecite{modecoupling} by replacing the summation in
Eq.~(\ref{boudaryeq}) with the operation of matrix multiplication.
Eq.~(\ref{boudaryeq}) can be written in terms of matrix as
\begin{subequations}
\label{operator}
\begin{eqnarray}
{\bf u}_{l}^{L} &= & \tilde {\bf u}_{l,n}(\omega, {\bf q}) + {\bf E}(-){\bf \Lambda}_{l}(-){\bf t}', \\
{\bf u}_{l}^{R}&=& {\bf E}(+){\bf \Lambda}_{l}(+){\bf t}''.
\end{eqnarray}
\end{subequations}
Here ${\bf E}(-)=\{{\bf e}_1, {\bf e}_2, \cdots, {\bf e}_{n'} \}$ is
the  matrix formed by the column eigenvectors for the reflected
normal modes.  ${\bf \Lambda}_{l}(-)$ is the diagonal matrix with
the diagonal terms ranging from $e^{i{\bf q}'_{1}\cdot {\bf R}_l},
e^{i{\bf q}'_{2}\cdot {\bf R}_l}, \cdots$ to $e^{i{\bf q}'_{n'}\cdot
{\bf R}_l}$. The column vector ${\bf t}'$ consists of the amplitude
reflection coefficients $t^{LL}_{n'n}$ with the different indices
$n'$.  Similar meaning holds for the notations ${\bf E}(+), {\bf
\Lambda}_{l}(+)$ and ${\bf t}''$ in the right lead, respectively. With the
help of Eq.~(\ref{operator}),  the propagation of the phonon
waves\cite{modecoupling,generalvalue,reviewus} from $l$ to $l+s$
unit cell can be described by the propagator ${\bf F}^s$ as ${\bf
u}_{l+s} = {\bf F}^s \cdot {\bf u}_{l}$, where ${\bf F}^s \equiv
{\bf E} \cdot {\bf \Lambda}_{s}\cdot {\bf E}^{(-1)}$. The inverse
matrix ${\bf E}^{(-1)}$ of the matrix ${\bf E}$  is pseudo-inverse
because ${\bf E}$ is not necessarily a square matrix.

Combining Eq.~(\ref{operator}) and the propagator ${\bf F}^s$ with
the equations of motion in the scattering region,  we can
solve\cite{reviewus} ${\bf u}_l^L$ and ${\bf u}_l^R$ in
Eq.~(\ref{boudaryeq}).  Then the amplitude transmission/reflection
coefficients ${\bf t}''$ and ${\bf t}'$ are gotten through the
formulas  ${\bf t}''={\bf \Lambda}_{l}^{-1}(+)\cdot {\bf
E}^{-1}(+)\cdot {\bf u}^R_l$ and ${\bf t}'={\bf
\Lambda}_{l}^{-1}(-)\cdot {\bf E}^{-1}(-)\cdot (\tilde {\bf
u}_{l,n}-{\bf u}^L_l)$.  The energy transmission
 $\mathcal{T}_n(\omega)$/reflection $\mathcal{R}_n(\omega)$ is
 related to the amplitude transmission/reflection\cite{reviewus,wangj06prb} as $\mathcal{T}_n(\omega)=\sum_{n''}
|t^{RL}_{n''n}|^2 \cdot {v}^R_{n''} / \tilde{v}^L_{n}$ and
$\mathcal{R}_n(\omega)=\sum_{n'} |t^{LL}_{n'n}|^2 \cdot {v}^L_{n'} /
{v}^L_{n}$, respectively, where ${v}^L$ and ${v}^R$ are the group velocities for
the left and the right.

The achiral carbon nanotube $(11,0)$ with the SW defect is optimized
with the second generation Brenner potential.\cite{dwbrenner} The
force constants are derived from the optimized structure under small
displacement. The phonon dispersion is calculated using the force
constants in the unit cell on the lead.
\begin{figure}[t]
\includegraphics[width=0.6\columnwidth]{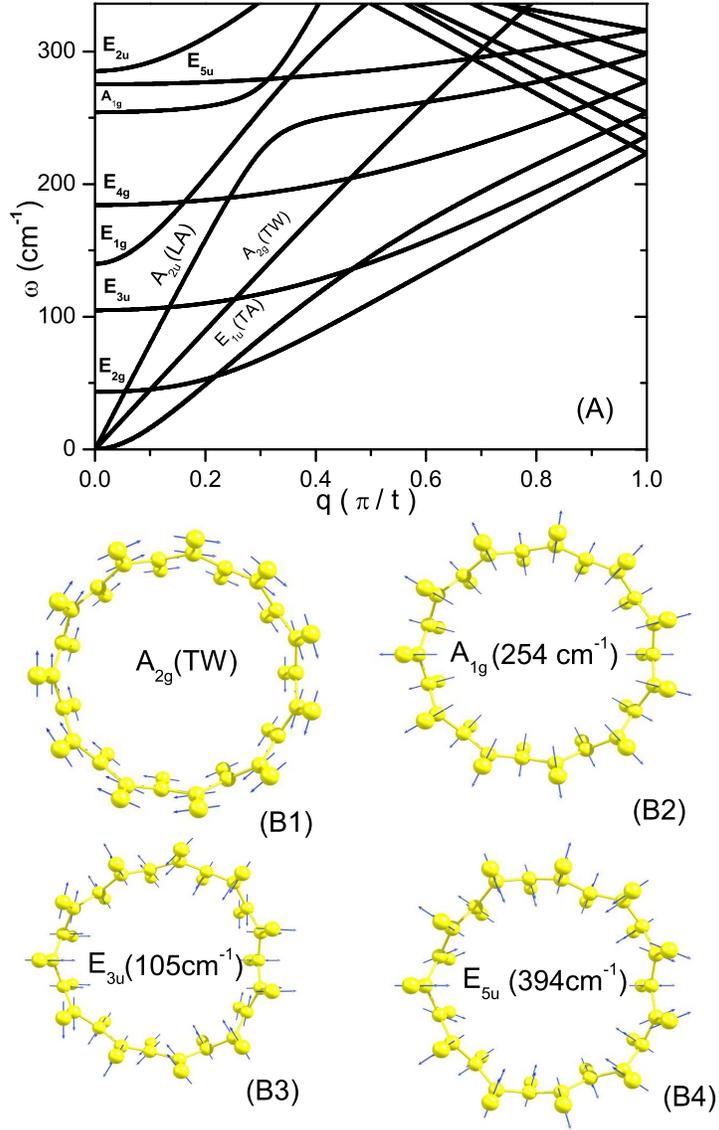}
\caption{\label{fig:phonondisp} Phonon dispersion and the symmetry
of phonon modes for the achiral $(11,0)$ CNT: \textbf{(A)} Phonon
dispersion and the symmetrical notations for each mode at $\Gamma$
points at low frequency. \textbf{(B1)-(B4)} Vibrations of the
$A_{2g}$, $A_{1g}$, $E_{3u}$ and $E_{5u}$ mode, respectively.
Frequency values indicated in the figure are at $\Gamma$ points. }
\end{figure}
 The calculated results are
shown in the picture (A) in Fig.~\ref{fig:phonondisp}. We further
analyze the symmetrical properties for each normal mode. The achiral
zig-zag carbon nanotube $(n,0)$ has the symmetry of the full line
group\cite{symvic,symbook} $L_Z={\bf T}_{2n}^1{\bf D}_{nh}$, where
${\bf T}_{2n}^1$ is the screw operation and ${\bf D}_{nh}$ is the
point group. However, it is not necessary to work with the full line
group. Instead, the point group is sufficient for the analysis of
phonon modes because the normal mode vectors at the $\Gamma$
point $(q=0)$ always transform as irreducible representations of the
isogonal point group. The point group isogonal to the line group,
\textsl{i.e.} with the same order of the principal rotational axis,
is ${\bf D}_{nh}$ for achiral tubes.  The isogonal point group for
the zigzag $(11,0) $ CNT is ${\bf D}_{22h}$. Each normal mode will
form a basis for an irreducible representation of the corresponding
line group. The symmetries for normal modes are found by decomposing
the representations of symmetrical operation into irreducible
representations.  The $12n$ branches of phonon modes for the zig-zag
$(n,0)$ CNT can be decomposed into the following irreducible
representations \cite{sysalon} at $\Gamma$ points
\begin{equation}
\label{operator1} \Gamma^{zig-zag}_{12n}=2(A_{1g}\oplus A_{2u}\oplus
B_{2g}\oplus B_{1u}) \oplus A_{2g}\oplus B_{1g}\oplus A_{1u}\oplus
B_{2u} \oplus\sum_{j=1}^{n-1}{3(E_{jg}\oplus E_{ju})}.
\end{equation}
We find that the calculated number of symmetries for $(11,0)$ CNT
from the force constants derived from Brenner potential strictly
respect the predicted isogonal ${\bf D}_{22h}$ symmetry. We list the point
group notations at $\Gamma$ points for the normal modes at low
frequency in the picture (A) in Fig.~\ref{fig:phonondisp}. It can be
seen from Fig.~\ref{fig:phonondisp} that symmetry properties for the
four acoustic branches are: the longitudinal acoustic mode (LA)
$\rightarrow A_{2u}$, the doubly degenerate transverse acoustic mode
(TA) $\rightarrow E_{1u}$, and the twist acoustic mode (TW)
$\rightarrow A_{2g}$. The calculated frequency at $\Gamma$ point for
the $A_{1g}$ radial breathing mode (RBM) is $254 \,\,{\rm cm}^{-1}$,
while the fitted value for RBM mode\cite{nanotubebook} is
$\omega_{RBM}(r)=\omega_{(10,10)}(r_{(10,10)}/r)^{1.0017\pm
0.007}=260\,\,{\rm cm}^{-1}$. Here $\omega_{(10,10)}$ and
$r_{(10,10)}$ are, respectively, the frequency and radius of the
$(10,10)$ armchair CNT, with values of $\omega_{(10,10)}=
165\,\,{\rm cm}^{-1}$ and $r_{(10,10)}=6.6785\,$\AA. The calculated
RBM frequency agrees with the predicted values. The calculated
vibrations for some typical normal modes are plotted in the
picture B in Fig.~\ref{fig:phonondisp}. It can be seen that the
phonon modes calculated from Brenner potential respect the
symmetries of CNTS well.

\begin{figure}[ht]
\includegraphics[width=0.5\columnwidth]{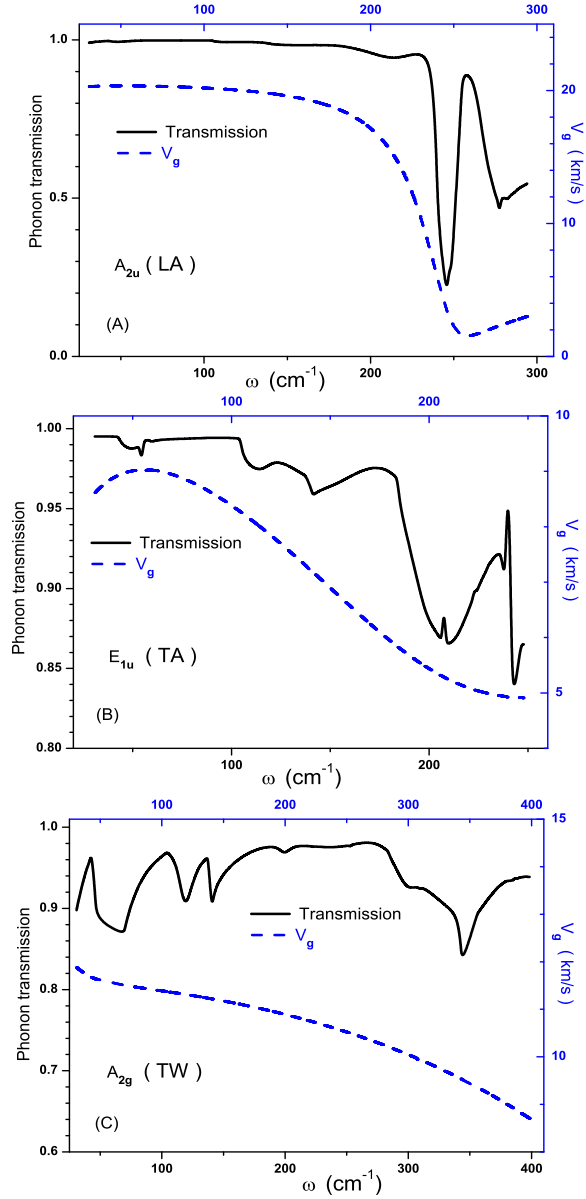}
\caption{\label{fig:acounstic} Frequency dependence of  phonon
transmissions and group velocities for acoustic modes: \textbf{(A)}
LA mode with the symmetry of $A_{2u}$; \textbf{(B)} TA mode with the
symmetry of $E_{1u}$;  \textbf{(C)} TW mode with the symmetry of
$A_{2g}$.
 }
\end{figure}

The calculated single mode phonon transmissions for the  LA, TA, and
TW acoustic branches are shown in Fig.~\ref{fig:acounstic} by the
solid lines. The group velocity for each mode is also plotted in
Fig.~\ref{fig:acounstic} with the dashed lines. It can be seen from Fig.~\ref{fig:acounstic} that the transmission for
each acoustic branch starts approximately at the value of one near the
$\Gamma$ point, which means that the SW defect CNTs pass long wave
acoustic phonons. With the increase of frequency, the group velocity
decreases for each acoustic mode. The transmissions for LA, TA, and
TW modes reduce in tendency. Especially, a gap in LA mode
transmission appears at frequency of $\omega=250\,\,{\rm cm}^{-1}$, where
the corresponding group velocity decreases to a very small value.
Compared with that of LA and TA modes, the transmission for TW mode
has more rich features. We think that this may be ascribed to the
higher symmetry of the atomic  vibrations for the TW mode as shown
in the picture (B1) in Fig.~\ref{fig:phonondisp}.

\begin{figure}[bt]
\includegraphics[width=0.8\columnwidth]{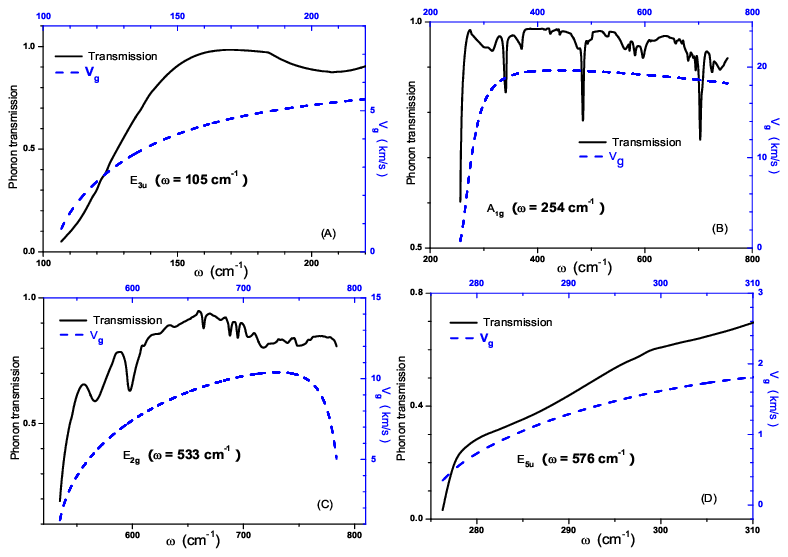}
\caption{\label{fig:optical} Frequency dependence of phonon
transmissions and group velocities for optical modes: \textbf{(A)}
$E_{3u}$ mode with $\omega=105 \, \,{\rm cm}^{-1}$ at $\Gamma$ point;
\textbf{(B)} $A_{1g}$ mode with $\omega=254 \, \,{\rm  cm}^{-1}$ at
$\Gamma$ point; \textbf{(C)} $E_{2g}$ mode with $\omega=533 \, \,
{\rm cm}^{-1}$ at $\Gamma$ point; \textbf{(D)} $E_{5u}$ mode with
$\omega=576 \, \, {\rm cm}^{-1}$ at $\Gamma$ point.
 }
\end{figure}

The transmissions as a function of frequency for optical modes are
shown in Fig.~\ref{fig:optical}.
In contrast to the transmission for acoustic modes, the optical mode
transmissions have very small values at $\Gamma$ points because of
the low group velocities. This is quite different from the
transmission for acoustic modes. The long wave optical phonons are
scattered strongly by the SW defect in CNTs.  Thus, the optical
phonons at $\Gamma$ points transport less energy in the SW defect
CNTs. Optical mode phonons near $\Gamma$ points contribute more to
local heating in the SW defect CNTs. We suggest that this feature
can be verified experimentally by measuring the intensities of
Raman-active modes such as $A_{1g}$. With the increase of frequency,
the optical mode transmissions increase as shown in
Fig.~\ref{fig:optical}. In comparison with the transmission of
$E_{3u}$ mode, the transmission for the RBM mode shows more peak
features. This can be explained by differences in their symmetrical
properties. As shown in the picture (B2), (B3) and (B4) in
Fig.~\ref{fig:phonondisp}, the vibrations of $A_{1g}$ mode have
higher symmetry than those of $E_{3u}$ mode while the vibrations of
the $E_{3u}$ and $E_{5u}$ modes seem to distort the symmetrical
structure of CNTs.   Therefore, the transmission for the $A_{1g}$ mode is more
sensitive to the SW defect in CNTs, which breaks the symmetry.
Together with the transmission features for the acoustic phonons
with higher symmetry such as the TW mode,  it can be inferred that the transmissions
both for the acoustic and optical phonons with high symmetrical properties are
sensitive to the symmetry breaking. This symmetrically
mode-dependent behavior for the single mode phonon transmission
maybe help to distinguish the different phonon modes through their
transport features after the symmetry breaking.

In summary, the single-mode phonon transmissions are calculated for
the symmetry broken CNT $(11,0)$ in the presence of the SW defect using the
mode-matching method. The symmetries of normal mode phonons are
analyzed in comparison with the phonon transmissions. It is found
that the acoustic phonon transmissions decrease with the reduced
group velocity at high frequencies while the optical phonon
transmissions have very small values near $\Gamma$ points. The
features of single-mode phonon transmission after symmetry breaking are related to the
symmetry property of each normal mode. These findings may help
to reveal the local heating problem in the defect CNTs. No evident
mode conversion has been observed during our calculation.  We also
calculated other different chirality CNTs in the presence of the SW
defect and similar findings can be made. Our present calculation
holds at moderately low temperature where anharmonic scattering does
not play an important role. When temperature is sufficiently high,
nonlinear scattering should be considered.

We thanks L\"{u} Jingtao  and Yang Huijie for helpful discussions.
This work was supported in part by a Faculty Research
Grant (R-144-000-173-101/112) of National University of Singapore. J.
Wang thanks National Natural Science Foundation of China (NSFC)
under the grant 10705023.



\begin{thebibliography}{01}

\bibitem{symvic} M. Damnjanovi\'{c}, I. Milo\v{s}evi\'{c}, T.
Vukovi\'{c} and R. Sredanoi\'{c}, Phys. Rev. B  \textbf{60},
2728(1999).

\bibitem{symbook} S. Reich, C. Thomsen, J. Maultzsch, \textsl{Carbon Nanotubes: Basic
Concepts and Physical Properties}, WILEY-VCH, 2004.

\bibitem{sysalon} O. E. Alon,  Phys. Rev. B \textbf{63}, 201403 (2001).

\bibitem{nanotubebook} R. Saito, G. Dresselhaus, and M. S.
Dresselhaus,   \textsl{Physical Properties of Carbons Nanotubes},
Imperial College Press, 1998.

\bibitem{nanotubereview} M. S. Dresselhaus, G. Dresselhaus,
P. Avouris, (Eds.),  \textsl{Carbon Nanotubes: Synthesis, Structure,
Properties and Application}, Topics Appl. Phys. \textbf{80} (2001).

\bibitem{ballistic1} P. Kim, L. Shi, A. Majumdar, and P. L. McEuen,
Phys. Rev. Lett. \textbf{87}, 215502 (2001).

\bibitem{ballistic2}H.-Y. Chiu, V. V. Deshpande, H. W. Ch. Postma, C. N. Lau, C. Mik\'{o}, L. Forr\'{o},
and M. Bockrath, Phys. Rev. Lett. \textbf{95}, 226101 (2005).

\bibitem{anderson} P. W. Anderson, Science, \textbf{177}, 393
(1972).

\bibitem{negfyamato}T. Yamamoto and K. Watanabe, Phys. Rev. Lett. \textbf{96},
255503 (2006).

\bibitem{phononcontrol} M. Terraneo, M. Peyrard, and G. Casati, Phys. Rev. Lett. \textbf{88}, 094302
(2002).

\bibitem{phononcontro2}C. W. Chang, D. Okawa, A. Majumdar, A. Zettl, Science \textbf{314},
1121 (2006).

\bibitem{phononcontro3}L. Wang and B. Li, Phys. Rev. Lett. \textbf{99}, 177208
(2007).

\bibitem{modecoupling} T. Ando, Phys. Rev. B  \textbf{44}, 8017
(1991).

\bibitem{generalvalue} P. A. Khomyakov, G. Brocks, Phys. Rev. B \textbf{70}, 195402
(2004).

\bibitem{reviewus} J.-S. Wang, J. Wang, and J. T. L\"{u},  Eur. Phys. J. B \textbf{62}, 381
(2008).

\bibitem{negfus} J.-S. Wang, J. Wang, N. Zeng, Phys. Rev. B \textbf{74}, 033408
(2006).

\bibitem{dwbrenner} D. Brenner, O. Shenderova, J. Harrison, S. Stuart,
 B. Ni,  and S. Sinnott, J. Phys.: Condens. Matter. \textbf{14}, 783 (2002).


\bibitem{ngfmingo} N. Mingo, D. A. Stewart, D. A. Broido, and D.
Srivastava, Phys. Rev. B \textbf{77}, 033418 (2008).

\bibitem{wangj06prb} J.~Wang, J.-S. Wang, Phys. Rev. B \textbf{74},
054303 (2006).


\end{thebibliography}
\end{document}